\documentstyle[12pt]{ioplppt}

\newcommand{\lco}{$\rm La_2CuO_4$}
\newcommand{\lsco}{$\rm La_{2-x}Sr_xCuO_4$}
\newcommand{\lesco}{$\rm La_{2-x-y}\-Eu_y\-Sr_x\-CuO_4$}
\newcommand{\leco}{$\rm La_{2-y}\-Eu_y\-CuO_4$}

\newcommand{\lgesco}{$\rm La_{1.99-x-y}Sr_xGd_{0.01}Eu_yCuO_4$}

\newcommand{\tn}{$T_N$}
\newcommand{\cuoc}{CuO$_6$}
\newcommand{\gape}{\mathrel{\mathop {_{_{\textstyle >}}} \atop ^{\textstyle \sim}}}
\newcommand{\co}{$\rm CuO_2$}
\newcommand{\Tl}{$T_{LT}$}
\newcommand{\Tn}{$T_{N}$}
\newcommand{\Eu}{$\rm Eu^{3+}$}
\newcommand{\gd}{$\rm Gd^{3+}$}

\begin{document}

\title{\raisebox{3cm}{\rightline{\underline{\it to be published
in Journal of Physics: Condensed Matter}}}
Strong enhancement of spin fluctuations in the
low-temperature-tetragonal phase of antiferromagnetically ordered
$\bf La_{2-x-y}\-Eu_y\-Sr_x\-CuO_4$}[Strong enhancement of spin
fluctuations in $La_{2-x-y}\-Eu_y\-Sr_x\-CuO_4$ ]

\author{V Kataev\dag\ddag\ftnote{3}{To
whom correspondence should be addressed. e--mail address:
kataev@ph2.uni-koeln.de}, A Validov\ddag\ , M H\"{u}cker\dag\ , H
Berg\dag\ and B B\"{u}chner\dag}

\address{\dag\ II. Physikalisches Institut, Universit\"{a}t zu K\"{o}ln,
50937 K\"{o}ln, Germany}

\address{\ddag\ Kazan Institute for Technical Physics, RAS, 420029
Kazan, Russia}

\begin{abstract}

Measurements of the static magnetization, susceptibility and ESR
of Gd spin probes have been performed to study the properties of
antiferromagnetically ordered \lesco\ $(x\leq 0.02)$ with the low
temperature tetragonal structure. According to the static magnetic
measurements the \co\ planes are magnetically decoupled in this
structural phase. The ESR study reveals strong  magnetic
fluctuations at the ESR frequency which are not present in the
orthorhombic phase. It is argued that this drastic enhancement of
the spin fluctuations is due to a considerable weakening of the
interlayer exchange and a pronounced influence of hole motion on
the antiferromagnetic properties of lightly hole doped \lco. No
evidence for the stripe phase formation at small hole doping is
obtained in the present study.

\end{abstract}

\maketitle

The structural phase transition from the low temperature
orthorhombic (LTO) phase to the low temperature tetragonal (LTT)
phase (hereafter the LT transition) in the rare-earth (RE) doped
\lsco\ high temperature superconductors \cite{Buchner,Craw} is
known to influence strongly the electronic properties of these
compounds at elevated levels of Sr (i.e. hole) doping. It is found
that the buckling of the \cuoc\ octahedra controls the occurrence
of superconductivity \cite{Buchner5} and leads to a static order
of spins and charges in form of stripes which has been observed in
Nd doped crystals \cite{Tranq,Zimmer}. In Eu doped \lsco\
anomalous slowing of spin dynamics \cite{ESRdynam} and
reappearance of magnetic order at low temperatures \cite{Wulfgar3}
for $x\ge 0.08$ have been found in ESR and $\mu$SR experiments,
respectively.

While spatially modulated spin correlations in \lsco\ are well
established at elevated hole concentrations, the situation is
controversial for the low hole doping regime. Neutron scattering
experiments do not reveal incommensurate spin correlations for
$x\le 0.05$ \cite{Yamada}. However, static magnetic measurements
and results from several local resonance techniques evidence the
formation of hole-free antiferromagnetic (AF) domains separated by
charged domain walls \cite{Cho1,Cho,Borsa,Suh}. The properties of
the ordered N\'eel state of the stripe phase in the doped cuprates
are intensively discussed theoretically \cite{Neto,Zaanen}.
Predictions of the stripe model agree well with the observed
suppression of the ordering temperature $T_N(x)$ with hole doping
\cite{Neto}. The effects of doped holes on the AF ordered \lco\
with LTO structure have been thoroughly studied by NQR
\cite{Cho,Borsa,Suh}. The experiments reveal magnetic fluctuations
in the ordered state at small hole doping which freeze out at low
$T$. It is argued that these fluctuations result from a collective
charge (stripe) motion in the planes thus giving evidence for a
stripe phase formation at very low hole doping \cite{Borsa,Suh}.
The relevance of this scenario to the underdoped \lco\ can be
verified by making a structural phase transition to the LTT phase
by RE doping \cite{Buchner,Craw}. One may anticipate a diminished
influence of holes on the spin dynamics in the LTT phase because
of the pinning of charge stripes which is known to occur in
compounds with large hole concentration \cite{Tranq}. However,
this expectation is not met in the recent NQR experiments on
\lesco\ which evidence a more dynamic behavior of Cu spins in the
LTT phase \cite{Suh2}.

To obtain an insight into the stripe phase formation and to study
the interplay between structure and magnetism in lightly hole
doped lanthanum copper oxide we have studied the magnetic
properties of Eu doped \lsco\ at low Sr concentrations by means of
static susceptibility, magnetization and electron spin resonance
(ESR) of Gd spin probes. We find striking changes of the magnetism
due to the LT transition: (i) the \gd\ ESR spectrum which exhibits
a partially resolved fine structure in both the paramagnetic and
the magnetically ordered state in the LTO phase, collapses into a
single narrow line for $x>0$ and $T<T_{LT}$, signalling strong
fluctuating magnetic fields in the hole doped samples at the ESR
frequency; (ii) a spin-flip transition in the magnetization curves
$M(H)$ clearly visible at $T>T_{LT}$ is strongly reduced for
$T<T_{LT}$ suggesting a considerable decrease of the interlayer
coupling in the LTT phase. We argue that the strong enhancement of
spin fluctuations in the hole doped samples is related to the
magnetic decoupling of the \co\ planes with LTT structure. Our
data give no indication for the relevance of charge stripes at
very low hole concentration.

Polycrystalline \lesco\ samples with $0\leq x \leq 0.03$ and
$0\leq y \leq 0.20$ were prepared as described elsewhere
\cite{Breuer}. To remove the excess oxygen the samples were
annealed in $\rm N_2$ atmosphere at $625 ^\circ C$ for 3 days.
The samples for ESR measurements were additionally doped by
0.5\% of Gd ions. The ESR experiments were carried out at 9.3
GHz. The static magnetic susceptibility $\chi(T)$ was measured
using a Faraday balance in a magnetic field of 10 kOe.
Measurements of the magnetization $M(H,T)$  in fields up to 140
kOe were performed with a vibrating sample magnetometer. The LT
transition temperatures \Tl\ $\sim 120-140$ K were verified by
X-ray diffraction.

In Fig.~\ref{esr1} we show a set of \gd\ ESR spectra (field
derivatives of absorption) of \lgesco\ samples. We find that in
the paramagnetic state, i.e. above the ordering temperature \tn,
the spectra of all samples are structured and most of the weight
of the spectra is below 2000~Oe (see e.g. spectra (1) in
Fig.~\ref{esr1}(a) and (b)). At low $T$ the shapes of the spectra
change. For the samples without Sr this concerns both the number
and the positions of the peaks. Nevertheless, the spectra remain
structured for both Eu free and Eu doped cases
(Fig.~\ref{esr1}(a), spectra (2) and (3)). However, doping of
\lco\ with {\em both} Eu and Sr leads to a qualitatively new and
striking effect: the ESR spectrum narrows into a single line at
$T<T_{LT}$ with a width less than the extent of the resolved
spectrum. Moreover, the line shifts considerably to higher fields.
We illustrate this in Fig.~\ref{esr1}(b) which shows a
representative set of spectra for the sample with $ x=0.008,\
y=0.20$. For $T>T_{LT}$ the resonance curves (Fig.~\ref{esr1}(b),
spectra (1) and (2)) resemble respective spectra for the Sr--doped
samples without Eu. However, below the LT transition the spectrum
acquires a shape close to a single Lorentzian line (see
Fig.~\ref{esr1}(b), spectrum (3)). Its spectral weight shifts to
much higher fields $(\gape 3000 Oe)$. The peak-to-peak width of
this line becomes smaller than the total extent of the structured
spectra. Eventually at low temperatures the structure in the
spectrum of the Eu doped sample partially reappears and the center
of weight of the spectrum shifts back to lower fields
(Fig.~\ref{esr1}(b), spectrum (4)). With increasing the Sr content
in \lesco\ the collapse of a structured ESR spectrum into a single
line occurs at progressively lower temperatures. The width of the
line gets smaller. The shaded areas in the T--x phase diagram in
Fig.~\ref{phase} map the regions where the collapsed ESR spectrum
is observed. We note that the collapse of the ESR spectrum does
not occur for the samples with the LTO structure (with and without
Eu) in the whole studied temperature range $(8K\leq T \leq 300K)$.
In particular, a respective spectrum for a hole doped sample
without Eu (i.e. with the LTO structure) shows a well resolved
structure (Fig.~\ref{esr1}(b), spectrum (5)).

A free \gd\ ion has a pure spin ground state $(S=7/2)$  with
8--fold degeneracy. This degeneracy is lifted in a magnetic field
which allows ESR transitions between the states $\mid\!S_z\!>$ and
$\mid\!S_z\pm 1\!>$. All of them require the same energy quantum
$g_{spin}\mu_B H_{res}=h\nu$. Therefore a single absorption line
is expected at the field $H_{res}\approx 3300 Oe$ for $\nu \approx
9.3GHz$ and a g--factor $g_{spin}\simeq 2$. However, the
crystalline electrical field (CEF) causes a small initial
splitting of \gd\ energy levels. As a result, resonance
transitions between different levels occur at different $H$ and
the respective spectrum consists of more than one line (so called
fine structure (FS) ESR spectrum) \cite{Abragam}. In \lco\ the
\gd\ ESR spectrum is rather complex \cite{Kat1,Rettori}. It
comprises a set of absorption lines which partially overlap. The
spectrum is highly asymmetric and its weight is considerably
shifted from $H\approx 3300 Oe$ to fields lower than $2000 Oe$.

\gd\ ESR studies of single crystals of \lsco\ $(x\leq 0.024)$ show
that in the AF state the spectrum additionally splits due to a
static internal field \cite{Rettori}. At $T\ll T_N$ the components
of the FS spectrum are well resolved only in the insulating case.
Presence of holes broadens the spectrum, nevertheless it remains
structured \cite{Rettori}. Qualitatively similar features are
observed in our experiments too. Also doping of insulating \lco\
by Eu does not change the behavior qualitatively. It causes only
an additional broadening and partial overlapping of individual
components of the FS spectra (Fig.~\ref{esr1}, spectrum (3)) due
to distortions of the CEF and due to a spatially inhomogeneous
\Eu\ Van Vleck magnetization. Thus one can conclude that the
striking collapse of the structured \gd\ ESR spectrum observed in
\lgesco\ is due to both the structural LT transition induced by Eu
doping {\em and} the presence of holes in the \co\ planes.

The narrowing of the \gd\ ESR spectrum in a single line resembles
the effect of exchange narrowing in magnetic resonance. This
similarity strongly suggests a {\em dynamic} cause for the
collapse of the \gd\ ESR spectrum. In our case the Gd spins are
highly diluted in the lattice and therefore Gd--Gd exchange is
apparently too small to be responsible for the narrowing of the
spectrum. However, studies of ESR in conventional metals give
evidence for a {\em relaxation} narrowing of the FS spectrum,
which is a single--ion process \cite{Barnes}. If the rate of a
certain ESR transition between FS split levels of a \gd\ ion is
strong enough to couple it with the neighboring transitions then
eventually the FS spectrum may collapse in a single narrow line.
In case of lightly Sr doped \lgesco\ the only thinkable source of
strong Gd spin relaxation is a fluctuating field produced by
copper spins. Therefore, from the ESR data we conclude that the LT
transition results in strong spin fluctuations in the \co\ planes
with frequencies of the order of the ESR frequency for some $x>0$
and $T$ (see Fig.~\ref{phase}).

To understand the drastic changes of the spin dynamics in the \co\
planes due to the LT transition we have measured the static
magnetic properties of \lesco. Representative susceptibility
curves $\chi(T)$ for the samples with $x=0;\ 0.010(1)$ and $y=0;\
0.2$ are shown in Fig.~\ref{sus}(a) and (b). Note that for the
Eu--doped samples a Van Vleck contribution of \Eu\ ions has been
subtracted from the raw susceptibility and magnetization data
\cite{Markus}. Furthermore, the data have been corrected for the
core diamagnetism $(\chi_{core}=-0.99\cdot 10^{-4} emu/mole)$ and
the Van-Vleck paramagnetism of the copper ions
$(\chi_{VV}=0.43\cdot 10^{-4} emu/mole)$ \cite{Allgeier} . The
$\chi(T)$ curves for $y=0$ and $0.2$ show a very similar behavior
for $T>T_{LT}$. There is a well defined N\'eel peak at the AF
ordering temperature \tn\ whose position shifts to lower $T$ with
increasing the Sr concentration $x$. The ordering temperatures as
a function of $x$ are plotted in Fig.~\ref{phase}. It is apparent
that \Tn\ of Eu doped samples follows the same dependence on  $x$
as is observed for \lsco\ (solid line in Fig.~\ref{phase}). Below
\Tl\ the N\'eel peak is not seen in the Eu doped compounds with
$T_N<T_{LT}$. However, according to recent $\mu$SR measurements
\cite{Wulfgar3} the \Tn\ values of these samples are similar to
the respective samples with the LTO structure as well. Remarkably,
the susceptibility of the Eu doped samples exhibits a pronounced
step--like increase which coincides with the LT transition
$(T_{LT}\approx 120-140K)$. The amplitude of the step is almost
the same for all samples with $x\leq 0.016$. For higher values of
$x$ the anomaly rapidly decreases and vanishes at $x=0.020$.

In Fig.~\ref{sus}(c) we show representative magnetization curves
$M(H)$ of Eu doped \lco\ acquired at temperatures slightly above
and below the LT transition. At $T>T_{LT}$ a step-like change of
$M(H)$ at $H_c\sim 40 kOe$ is seen similar to that observed in
pure \lco\ crystals \cite{Thio}. Remarkably, below \Tl\ the
step--like change of $M(H)$ is strongly reduced. Similar changes
of $M(H)$ due to the LT transition are apparent also in Sr doped
samples with $T_N>T_{LT}$.

In the \co\ planes with the LTO structure the spins are slightly
canted out of the planes because of the tilting of the \cuoc\
octahedra and the Dzyaloshinsky--Moriya (DM) antisymmetric
anisotropic exchange. Therefore a net out-of-plane ferromagnetic
moment (DM moment) arises in every layer \cite{Thio}. In the LTT
structure the tilting of the Cu-O-Cu bonds is still present in one
crystallographic direction \cite{Buchner,Craw} and allows for the
DM interaction too \cite{Coffey}. In principle, two kinds of spin
arrangements are possible in the latter structure: one with and
another without the DM moments \cite{Coffey}. Which of these two
configurations should be the ground state of the LTT phase of
\lco\ is still a question of an intense theoretical discussion
(see e.g. \cite{Kosh,Vier,Stein}). In the following we show that
our static magnetic data give strong evidence that the DM moments
are present in the LTT phase of Eu doped \lsco.

It is well established that in the LTO phase, below \Tn, the DM
moments are ordered antiferromagnetically due to the interplane
coupling of the Cu spins \cite{Thio}. Because the DM moments
produce an additional contribution $\chi_{DM}$ to the total
susceptibility, the measured $\chi$ in the LTO phase is
significantly larger than that expected for an isotropic
two-dimensional $S=1/2$ Heisenberg antiferromagnet $\chi_{2DHAF}$
(see the respective curves in Fig.\ref{sus} and
Ref.\cite{Johnston91}). Our experimental data clearly show that at
the LT structural transition the difference between
$\chi_{2DHAF}$ and the measured susceptibility of \lesco\ at 10
kOe {\em even increases} (see Fig.\ref{sus}a,b), thus suggesting
that the DM moments are present also in the LTT phase. The same
conclusion emerges from the analysis of the field dependence of
the magnetization $M(H)$. As one can see in Fig.~\ref{sus}(c), at
all fields it is larger than that of a 2DHAF by the amount
$M_{DM}$ which is the out-of-plane magnetization due to the DM
moments. The main feature of the magnetization curves $M(H)$ for
$T>T_{LT}$, the step--like increase by $M_{SF}$, is related to the
spin--flip (SF) of these moments at the critical field $H_c$
\cite{Thio}. Above $H_c$, due to the spin--flip, $M_{DM}$ makes
the dominant contribution to the measured $M(H)$. Remarkably, the
value of $M_{SF}$  which measures the staggered part of the DM
magnetization, drops down at the LT transition (see inset of
Fig.\ref{sus}c), and the critical field $H_c$ is not defined for
the LTT phase \cite{remark}. Nevertheless, at fields $H>100 kOe$
$M(H)$-curves measured slightly above and below the LT transition
come very close together, i.e. the high field magnetization which
is mostly due to the DM exchange obviously does not change at the
LT transition. This is also evident from the T-dependence of
$M_{DM}$ measured at $H=140kOe$ (see inset of Fig.\ref{sus}c).
Thus, one may conclude that \co\ planes with the LTT structure
carry the same net DM moments as in the LTO phase. Hence, the
strong reduction of the spin--flip transition evidences the
considerable weakening of the interlayer exchange in the LTT phase
which is responsible for the AF arrangement of the DM moments.

The reason for such a dramatic effect of the structural transition
on the magnetic coupling of the \co\ planes has been suggested in
Ref.\cite{Shamoto} where the neutron diffraction (ND) studies of
$\rm La_{2-y}Nd_yCuO_4$ crystals with the LTT structure are
presented. For the LTT phase these studies reveal spin structures
with a weakened and frustrated interlayer exchange which leads to
the coexistence of magnetic domains with effective AF and
ferromagnetic interlayer coupling. In accordance with the ND
results our static magnetic measurements evidence that the
magnetic decoupling of the \co\ planes due to such a frustration
takes place in insulating as well as in hole doped \leco\ and
results macroscopically in the absence of the critical spin-flip
field and a considerable increase of the susceptibility and
magnetization at low fields.

It is straightforward to relate a strong decoupling of the layers
due to the LT transition with the strong magnetic fluctuations
observed in the hole doped \lesco\ samples at ESR frequencies.
Magnetic fluctuations in the AF ordered {\em hole doped} \lco\
with the LTO structure were observed by NQR in the MHz--frequency
window \cite{Cho,Borsa,Suh}. Our ESR data suggest the presence of
strong magnetic fluctuations in hole doped \lesco\ with the LTT
structure on a much higher frequency scale. A possible reason for
such a drastic enhancement of the magnetic dynamics is an
effective reduction of the dimensionality of the AF correlations
in the \co\ planes due to interlayer decoupling. As a result the
magnetic correlations should be less stable against perturbations
produced by moving holes. Consequently the frequency of the spin
fluctuations may be considerably enhanced compared to the LTO
phase. In a certain temperature range $\Delta T<T_{LT}$  it may
approach the ESR frequency window and thus provide a channel for a
strong spin relaxation of Gd probe. Thus our data strongly suggest
that magnetic dynamics due to the charge motion is not inhibited
in the LTT structure at small hole doping.

The NQR data on the LTO \lsco\ samples were explained assuming
that holes form stripes or loops in the layers \cite{Borsa,Suh}.
It is argued that the motion of these charge objects serving as
anti--phase boundaries for the hole free magnetic domains causes
flips of the Cu spins. This motion freezes out at low $T$ leading
to a recovery of the sublattice magnetization and to a peak in the
$T$--dependence of the $\rm ^{139}La$ spin--lattice relaxation
rate $(1/T_1)^{La}$ at $T_f\sim 10-16$ K \cite{Cho,Borsa,Suh}. In
this scenario changing the structure from LTO to LTT should lead
to weakening of magnetic fluctuations in the AF ordered \co\
planes since charge stripes are pinned in the LTT phase
\cite{Tranq}.

This expectation is in a striking contrast with our experimental
observations. For a qualitative understanding of our data an
assumption of collective charge motion is even not necessary. In
particular, we find no evidence for a pinning of charge stripes
which is known to exist in the LTT phase at higher levels of
hole doping. Remarkably in a recent $\rm ^{139}La$ NQR study of
lightly Sr doped \lesco\ $(x=0.010; 0.015)$ \cite{Suh2} the peak
in $(1/T_1)^{La}$ was found to occur at a considerably lower
temperature ($T_f\sim 6$ K) than in compounds with LTO structure
\cite{Cho,Borsa,Suh}. As this peak signals freezing of the spin
fluctuations on the NQR time scale, its shift to lower $T$ is
also suggestive for a more dynamic magnetic behavior of the LTT
phase. Therefore for a better understanding of the spin dynamics
it would be desirable to perform a systematic NQR study of
lightly hole doped \lesco.

In summary, we have measured susceptibility, magnetization and ESR
of Gd spin probes in order to study the magnetic properties of the
LTT phase of \lesco\ with small hole doping $x\leq 0.03$. The
static magnetic measurements evidence a drastic reduction of the
interlayer coupling in the LTT phase. In the ESR measurements of
the hole doped samples we observe a collapse of the Gd spectrum
into a single narrow line at $T<T_{LT}$ which signals strong
magnetic fluctuations at the ESR frequency. We conclude from the
data that the distinct differences in the properties of the AF
ordered LTT phase of \lesco\ at low levels of hole doping arise
due to the effective interlayer magnetic decoupling, which
enhances the influence of mobile holes on the antiferromagnetism
of this compound. In particular, the data give no evidence for
charge stripes at small hole concentrations.

This work was supported by the DFG through SFB 341 and by NATO
CR grant No. 972046. The work of V.K. and A.V. was supported in
part by the Russian State HTSC Program (project No. 98001) and
by the RFBR (project No. 98-02-16582). M.H. acknowledges support
by the Graduiertenstipendium des Landes Nordrhein-Westfalen.

\Figures

\begin{figure}
\caption[]{(a) -- \gd\ ESR spectra of insulating $\rm
La_{1.99-y}\-Eu_{y}\-
           Gd_{0.01}\-CuO_{4}$. Curves (1) and (2) correspond to
           the sample without Eu in the paramagnetic state and
           in the AF state at 10K. Curve (3) at 10K corresponds
           to the sample doped by Eu $(y=0.15)$;\\
           (b) -- \gd\ ESR spectra of $\rm La_{1.99-x-y}\-Sr_{x}\-Eu_{y}
           Gd_{0.01}\-CuO_{4}$. Curves (1)--(4) show the
           evolution of the spectrum of the sample with
           $x=0.008$ and $y=0.20$ with decreasing temperature.
           Note the collapse of the spectrum into a nearly
           single line below the LT transition. The dotted line
           represents a Lorentzian fit. In the bottom panel a
           respective spectrum (5) of the sample without Eu is
           shown for comparison.}
\label{esr1}
\end{figure}

\begin{figure}
\caption[]{Static magnetic susceptibility $\chi (T)$ at $H=10$
           kOe (a), (b) and magnetization $M(H)$ (c) of $\rm
           La_{2-x-y}\-Eu_{y}\-Sr_{x}\-CuO_{4}$.
           Solid lines and triangles are predictions for the $S=1/2$ 2DHAF by
           the Schwinger boson mean field theory \cite{Auerbach}
           and by Monte Carlo simulation \cite{Odabe}, respectively,
           with an in-plane exchange $J\sim 1400K$.
           $\chi_{DM}$ and $M_{DM}$ denote contributions due to DM exchange.
           $M_{SF}$ is the spin-flip magnetization.
           The inset shows the T-dependences of $M_{DM}$ at $H=140 kOe$
           and of $M_{SF}$. (for details see text)}

\label{sus}
\end{figure}

\begin{figure}
\caption[]{Structural (\Tl) and magnetic (\Tn) phase diagram of
           $\rm La_{1.8-x}Sr_{x}Eu_{0.2} CuO_{4}$. The
           $T_{LT}(x)$--phase boundary is obtained by X-ray
           diffraction (dotted line). Solid circles are the \tn\
           values obtained from the susceptibility data. The
           solid line shows the $T_N(x)$--dependence for $\rm
           La_{2-x} Sr_xCuO_{4}$. The shaded ellipses are the
           regions where the \gd\ ESR spectra of the samples with
           $x=0.008; 0.014; 0.015; 0.017$ and $0.020$ collapse
           into a single line. The collapse is not observed for
           the other two Eu--doped samples studied in this work
           ($x=0.0$ and $0.03$), as well as for all samples
           without Eu.}
\label{phase}
\end{figure}

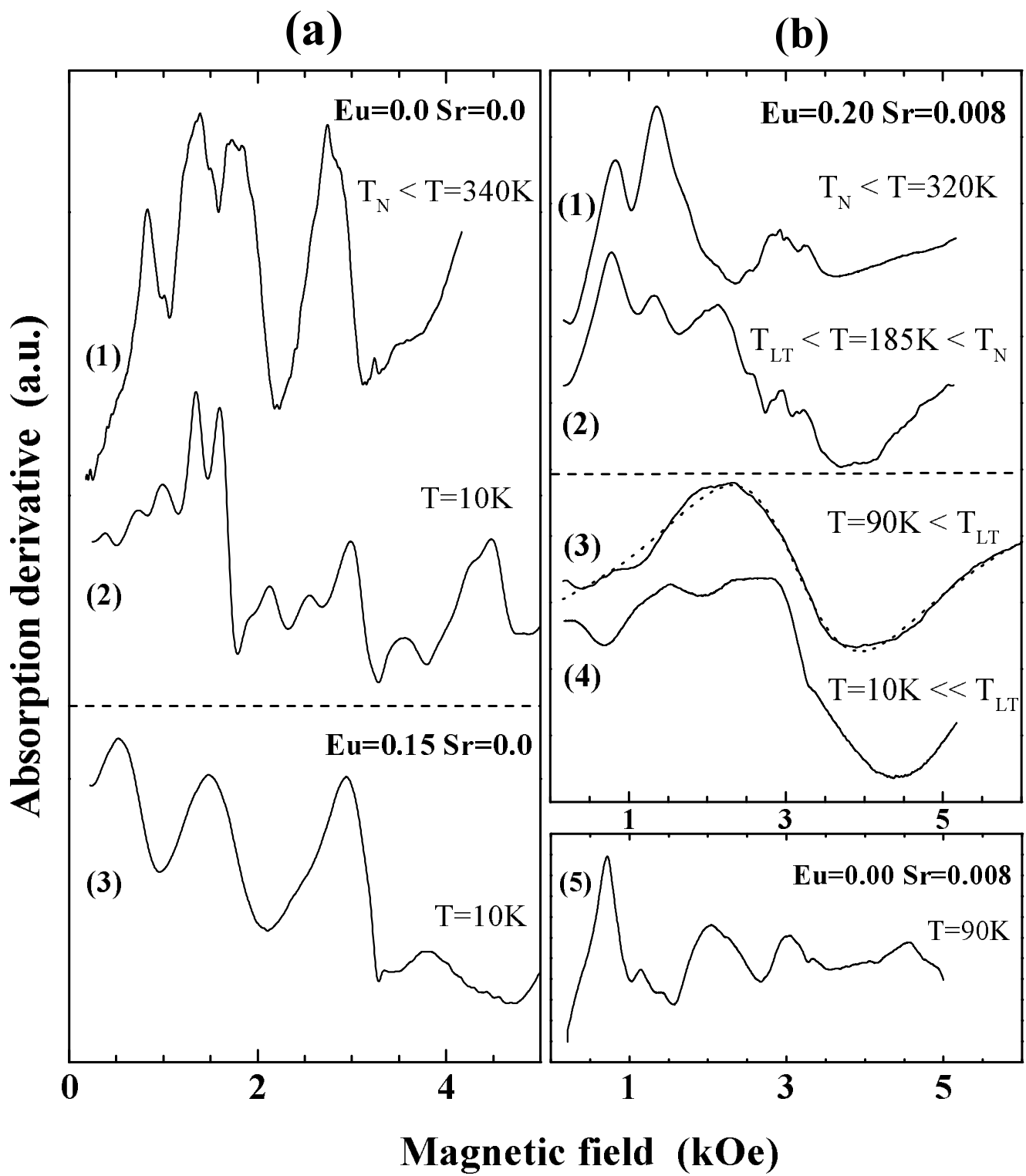
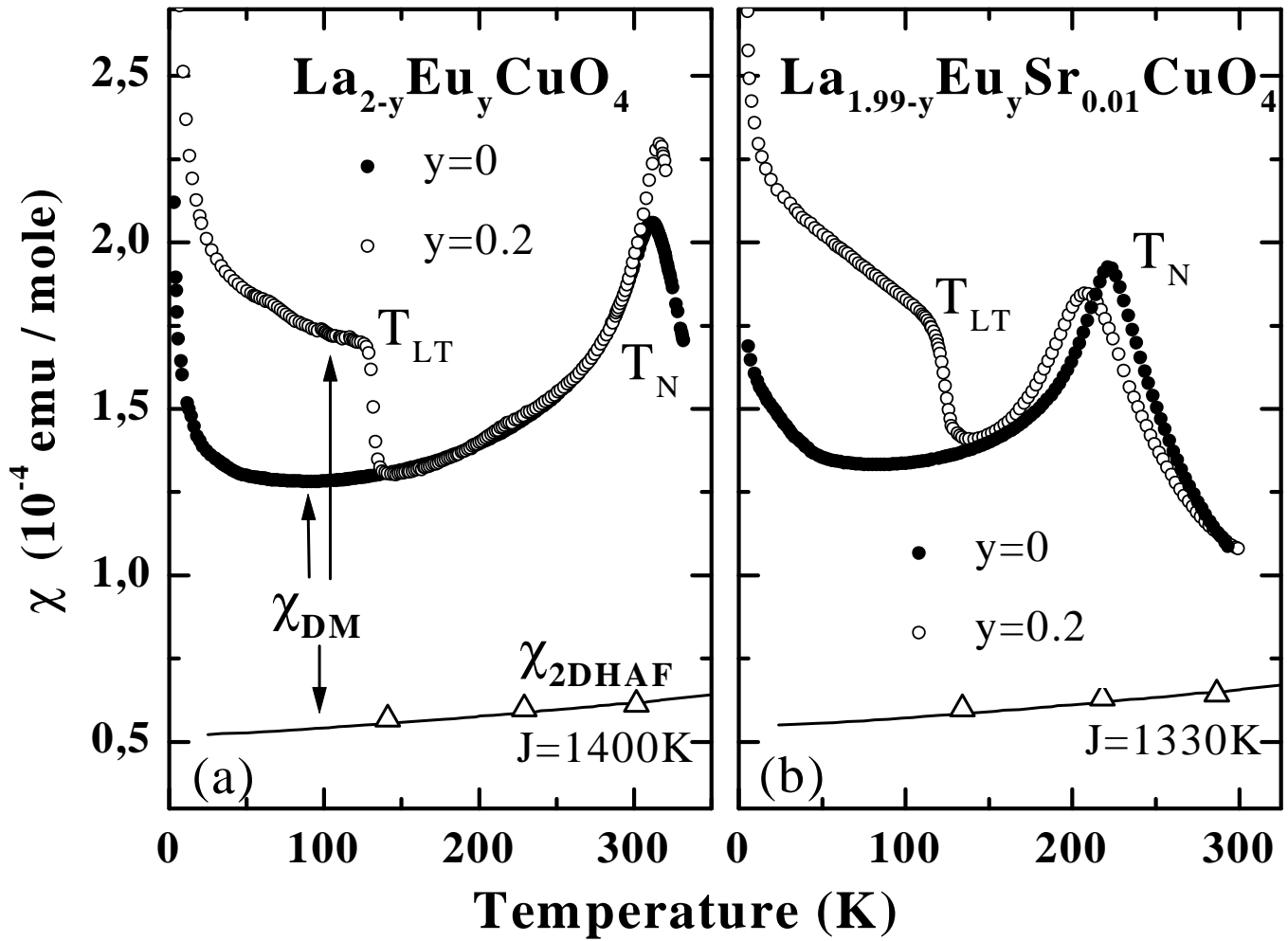
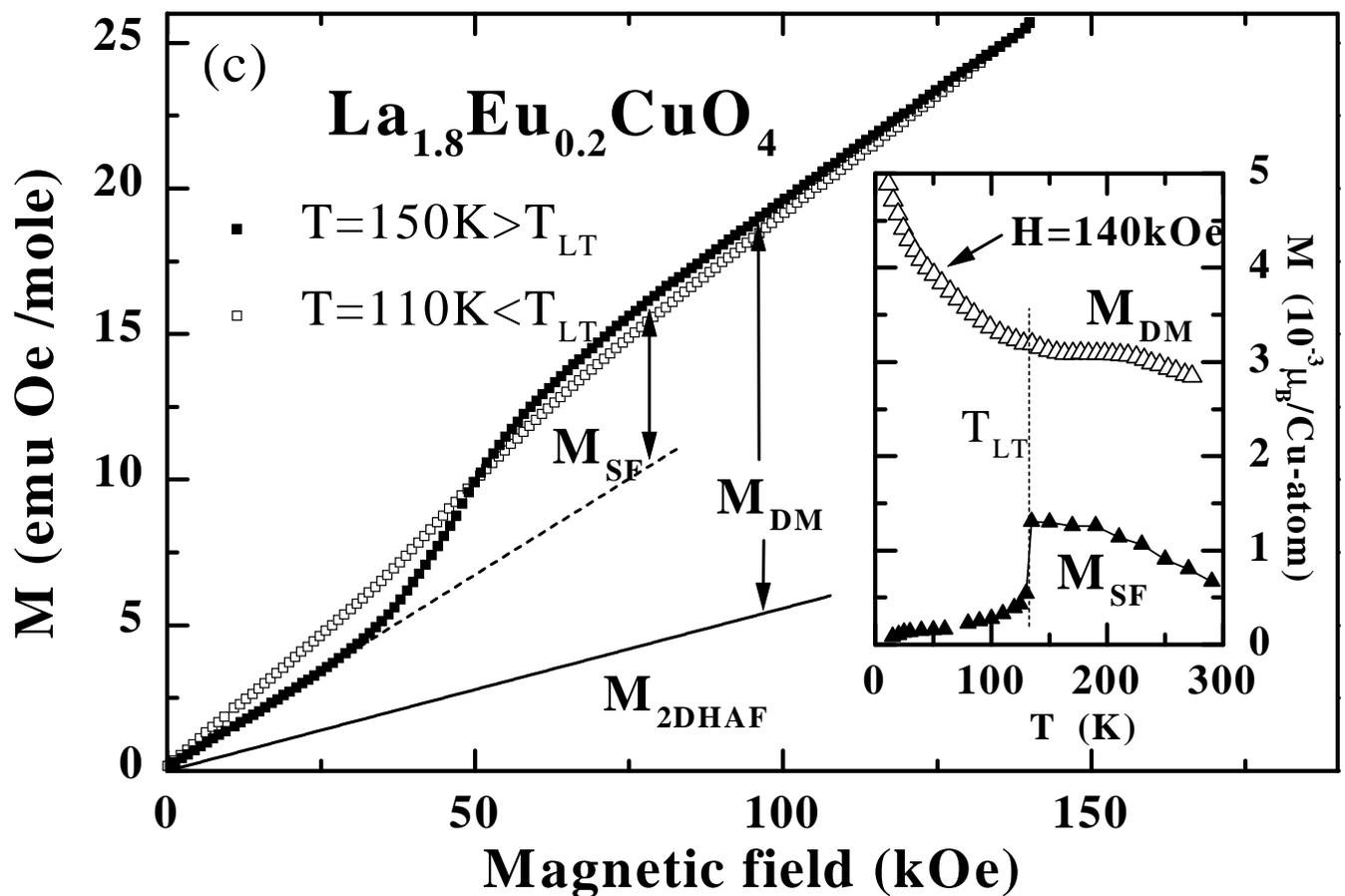
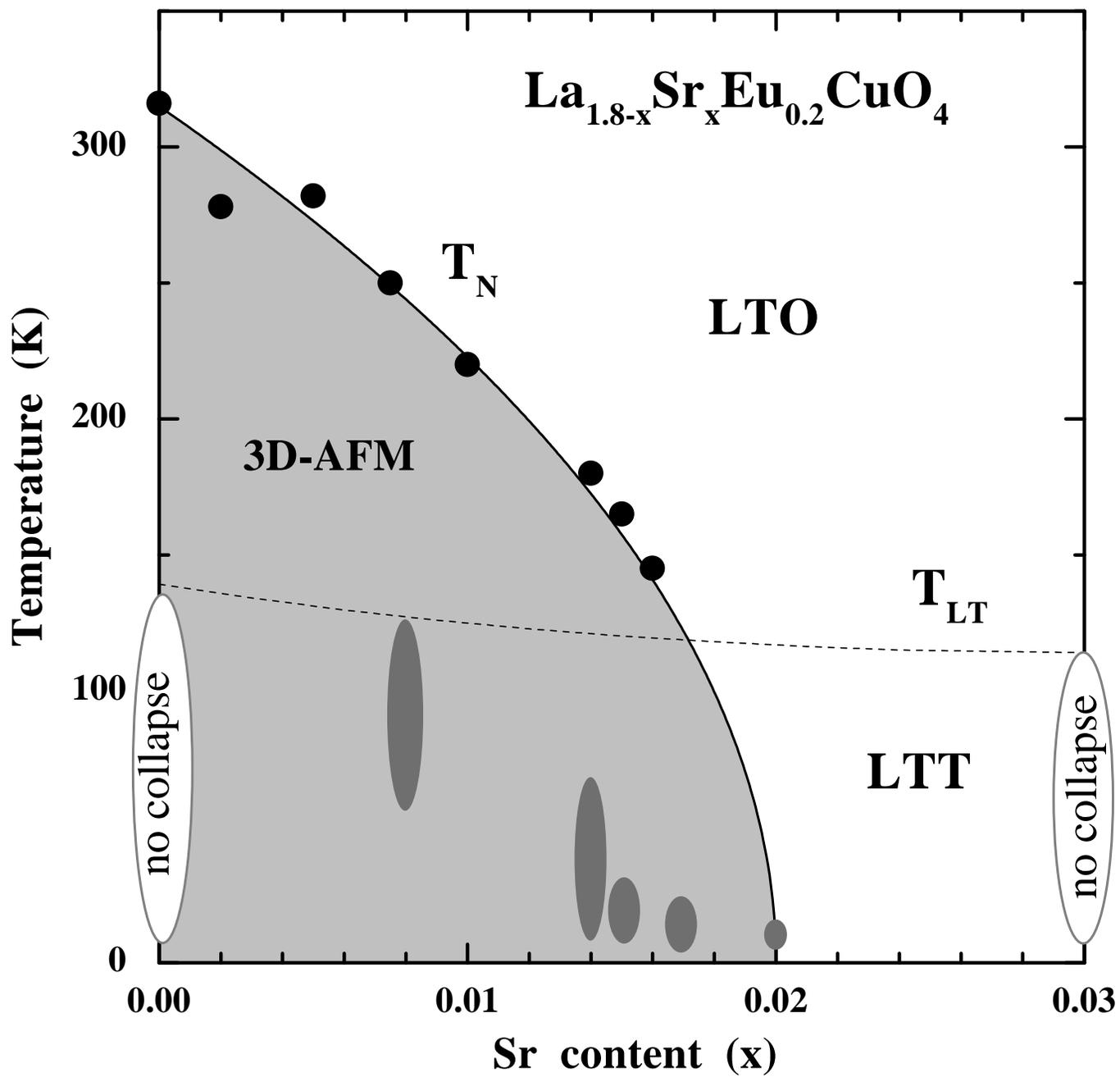
\end{document}